\documentclass[sigplan,screen]{acmart}
\usepackage{algorithm,algpseudocode}
\AtBeginDocument{%
  \providecommand\BibTeX{{%
    \normalfont B\kern-0.5em{\scshape i\kern-0.25em b}\kern-0.8em\TeX}}}

\copyrightyear{2022} 
\acmYear{2022} 
\setcopyright{acmcopyright}
\acmConference[ESSE 2022]{2022 The 3rd European Symposium on Software Engineering}{October 27--29, 2022}{Italy}
\acmBooktitle{2022 The 3rd European Symposium on Software Engineering (ESSE 2022), October 27--29, 2022, Rome, Italy}
\acmPrice{15.00}
\acmDOI{10.1145/3571697.3571705}
\acmISBN{978-1-4503-9730-8/22/10}

\begin{document}

%%
%% The "title" command has an optional parameter,
%% allowing the author to define a "short title" to be used in page headers.
\title{URegM: a unified prediction model of resource consumption for refactoring software smells in open source cloud}

%%
%% The "author" command and its associated commands are used to define
%% the authors and their affiliations.
%% Of note is the shared affiliation of the first two authors, and the
%% "authornote" and "authornotemark" commands
%% used to denote shared contribution to the research.
\author{Asif Imran}

\email{aimran@csusm.edu}
\orcid{0000-0002-1780-0296}

\affiliation{%
  \institution{California State University San Marcos}
  \streetaddress{333 S Twin Oaks Valley}
  \city{San Marcos}
  \state{California}
  \country{USA}
  \postcode{92096}
}

\author{Tevfik Kosar}

\email{tkosar@buffalo.edu}
\orcid{0000-0002-1780-0296}

\affiliation{%
  \institution{University at Buffalo}
  \streetaddress{Davis Hall}
  \city{Buffalo}
  \state{New York}
  \country{USA}
  \postcode{14260}
}
%%
%% By default, the full list of authors will be used in the page
%% headers. Often, this list is too long, and will overlap
%% other information printed in the page headers. This command allows
%% the author to define a more concise list
%% of authors' names for this purpose.
%\renewcommand{\shortauthors}{Trovato and Tobin, et al.}

%%
%% The abstract is a short summary of the work to be presented in the
%% article.
\begin{abstract}
The low cost and rapid provisioning capabilities have made the cloud a desirable platform to launch complex scientific applications. However, resource utilization optimization is a significant challenge for cloud service providers, since the earlier focus is provided on optimizing resources for the applications that run on the cloud, with a low emphasis being provided on optimizing resource utilization of the cloud computing internal processes. Code refactoring has been associated with improving the maintenance and understanding of software code. However, analyzing the impact of the refactoring source code of the cloud and studying its impact on cloud resource usage require further analysis. In this paper, we propose a framework called \textit{Unified Regression Modelling (URegM)} which predicts the impact of code smell refactoring on cloud resource usage. We test our experiments in a real-life cloud environment using a complex scientific application as a workload. Results show that \textit{URegM} is capable of accurately predicting resource consumption due to code smell refactoring. This will permit cloud service providers with advanced knowledge about the impact of refactoring code smells on resource consumption, thus allowing them to plan their resource provisioning and code refactoring more effectively.
\end{abstract}

%%
%% The code below is generated by the tool at http://dl.acm.org/ccs.cfm.
%% Please copy and paste the code instead of the example below.
%%
\begin{CCSXML}
<ccs2012>
   <concept>
       <concept_id>10011007.10011074.10011111.10011696</concept_id>
       <concept_desc>Software and its engineering~Maintaining software</concept_desc>
       <concept_significance>500</concept_significance>
       </concept>
 </ccs2012>
\end{CCSXML}

\ccsdesc[500]{Software and its engineering~Maintaining software}

%%
%% Keywords. The author(s) should pick words that accurately describe
%% the work being presented. Separate the keywords with commas.
\keywords{resource usage prediction, scientific application in cloud, unified regression modelling}

%% A "teaser" image appears between the author and affiliation
%% information and the body of the document, and typically spans the
%% page.

%%
%% This command processes the author and affiliation and title
%% information and builds the first part of the formatted document.
\maketitle

\section{Introduction}
Cloud computing is the dynamic provisioning of resources from a shared resource pool, which can be provisioned by a pay-as-you go mechanism \cite{armbrust2009above}. The dynamic nature and rapid provisioning of resources have made the cloud a desirable platform to run distributed scientific applications which are resource intensive. However, cloud service providers are constantly aiming to optimize resource consumption at the system end to make more resources available to its customers. Recent studies have shown that positive relationship exists between code smells and resource usage \cite{imran2021impact}. However, predicting this change of resource usage due to refactoring of code smells present in source code of the cloud have not been conducted. In this paper, we propose a model called \textit{Unified Regression Model (URegM)} that predicts the impact of code smell refactoring on resource consumption of cloud computing processes. To the best of our knowledge, relationship between code smell refactoring of cloud source code and establishing its relationship to resource consumption at the data center level have been conducted to a limited extent. 

In the cloud environment, optimizing resource usage of the cloud's own processes will provide business edge to the cloud service providers as they can provision the excess resource to perform the tasks of the cloud customers. However, the impact needs to be tested by running real life applications in the cloud. In this paper, we use a real-life scientific application as a workload to test the proposed framework. The selected scientific application is designed to perform in distributed environment and it can harness the power of the cloud to complete the jobs. The tasks in this application is divided in groups and distributed across multiple virtual machine (vm) instances of the cloud for parallel and collaborative computing. As a result, using the scientific application as a workload to test the resource usage of the cloud will provide an ideal and real life scenario to analyze performance. 

Earlier techniques primarily focus on predicting resource provisioning for running scientific applications in cloud. However, effective framework is required that will provide advanced knowledge on the impact of refactoring code smells on resource consumption as this will enable cloud service providers to utilize the extra resource saved by refactoring. Alternatively, if certain refactoring increases resource consumption, this prediction framework will update the cloud service provider about it. Code smells have been traditionally refactored to improve maintainability and understandability of source codes. Exiting studies focus on extensively testing software code to identify bugs that may result in code smells \cite{raksawat2021software}. However, code smells can be refactored in cloud to optimize resource usage by the cloud computing processes. This extra resource can be provisioned to support the resource requirement of resource intensive scientific applications. To the best of our knowledge, existing methods cannot accurately predict the impact of refactoring code smells on resource consumption in cloud environment.

To address the above problem, a novel prediction approach to analyze impact of code smell refactoring on resource usage is proposed which focuses on selection of features using Genetic Algorithm (GA) and unifies 4 regression algorithms to improve performance. The framework is tested using a real-life workload called \textit{WildfireDB} which is a scientific application designed to simulate wildfire situations using large volume of historical and vegetation data. This tool has simulation components to analyze wildfire situations to be studied by climate change ecologists. The main contributions of this paper are provided below:

\begin{itemize}
    \item \textit{URegM} is a framework to predict the change in resource consumption due to code smell refactoring of cloud source code. The performance of the framework is tested for running real life scientific applications in cloud.
    \item A dataset is used with selection of correct features of the cloud platform such as CPU, memory, weighted methods per class, lookahead which is used for training \textit{URegM} model. 
    \item Comparative analysis of \textit{URegM} to existing models like REAP \cite{senturk2018resource} show that the proposed model outperforms the current models in terms of \textit{mse, rmse, accuracy,} and \textit{execution time}.
\end{itemize}

Rest of the paper proceeds as follows. Section \ref{relatedwork} identifies the related research conducted in the field of predicting resource usage and highlights the importance of optimizing cloud resource consumption. Section \ref{uregm} provides details of the proposed \textit{URegM} framework and the prediction approach. Section \ref{experiment} highlights the setup of experimental environment and analyzes obtained results. Section \ref{conclusion} concludes the paper and identifies scope of future research.
\section{Related Work}
\label{relatedwork}

Park et al. investigated whether existing refactoring techniques support resource-efficient software creation or not \cite{park2014investigation}. Since resource efficient software was critical in mobile environments, they focused their study on mobile applications. Results showed that specific refactoring techniques like \textit{Extract Class} and \textit{Extract Method} worsened energy consumption because they did not consider power consumption in their refactoring process. The goal was to analyze the resource efficiency of the refactoring techniques themselves, and they stated the need for resource-efficient refactoring mechanisms for code smells. Imran et al. conducted a study on the impact of code smell refactoring on resource usage \cite{DBLP:conf/seke/ImranK20}. They used automatic refactoring tool to detect and remove code smells, followed by calculating resource consumption change for a specific workload. However, they did not apply machine learning to predict the impact. 

Platform-specific code smells in High-Performance Computing (HPC) applications were determined by Wang et al. \cite{wang2014platform}. AST-based matching was used to determine smells present in HPC software. The authors claimed that the removal of such smells would increase the speedup of the software. The assumption was that specific code blocks performed well in terms of speedup in a given platform. Perez-Castillo et al. stated that excessive message traffic derived from refactoring god class increased a system’s power consumption \cite{perez2014analyzing}. It was observed that power consumption increased by 1.91\% (message traffic = 5.26\%) and 1.64\% (message traffic = 22.27\%), respectively, for the two applications they analyzed. The heavy message-passing traffic increased the processor usage, which proved to be in line with the increase in the power consumption during the execution of those two applications. 

An automatic refactoring tool that applied the \textit{Extract Class} module to divide \textit{god class} smell into smaller cohesive classes was proposed in \cite{fokaefs2011jdeodorant}. The tool aimed to improve code design by ensuring no classes were large enough, which was challenging to maintain and contained a lot of responsibilities. The tool refactored code by suggesting \textit{Extract Class} modifications to the users through a User Interface. The tool was incorporated into the Eclipse IDE via a plugin. The authors consulted an expert in the software quality assessment field who gave his expert opinion to identify the effectiveness of the tool. Despite the existing effort in refactoring code smells, however, impact prediction of code smell refactoring on cloud resource consumption had been analyzed to a limited extent.

Liu et al. \cite{liu2017adaptive} enabled automatic resource maintenance in a cloud environment via forecasting the workload in advance. The cluster trace data from Google was used as a dataset in this regard. Ghobaei-Arani et al. \cite{ghobaei2018autonomic} used reinforcement learning to predict future resource requirements based on the current workload on the cloud. Chen et al. \cite{chen2015self} proposed a \textit{Fuzzy Neural Network (FNN)} to analyze current resource demand in the cloud and predicted future resource requirements. Shaw et al. \cite{shaw2019energy} proposed a novel predictive anti-correlated placement algorithm which made the cloud energy efficient by managing resources effectively.  We saw that most of the prediction frameworks for the cloud were focused on effective resource provisioning for tasks that were executed on the cloud, however, research needed to be done to analyze and predict resource usage requirements for the cloud environment itself.
\section{Resource prediction: \textit{URegM} method}
\label{uregm}

In this section, we present a case study where we amalgamate the four regression models to improve performance in terms of \textit{mse, rmse, accuracy,} and \textit{execution time}. We base our case study on one scientific application discussed later in this paper. The cloud environment has been specified to be\textit{ OpenStack}. Since cloud computing has become a significant paradigm where many real-life applications are executed, we select the open source cloud computing platform for our study. Cloud also offers a ubiquitous and parallel infrastructure to run critical scientific applications which are resource-intensive. Predicting the impact of cloud resource requirements after refactoring code smells of the cloud platform itself will be a significant contribution. This will enable cloud service providers to improve the resource scheduling scheme of the cloud. 

\begin{figure*}
    \centering
    \setlength{\belowcaptionskip}{-12pt}
        \includegraphics[width=1\textwidth]{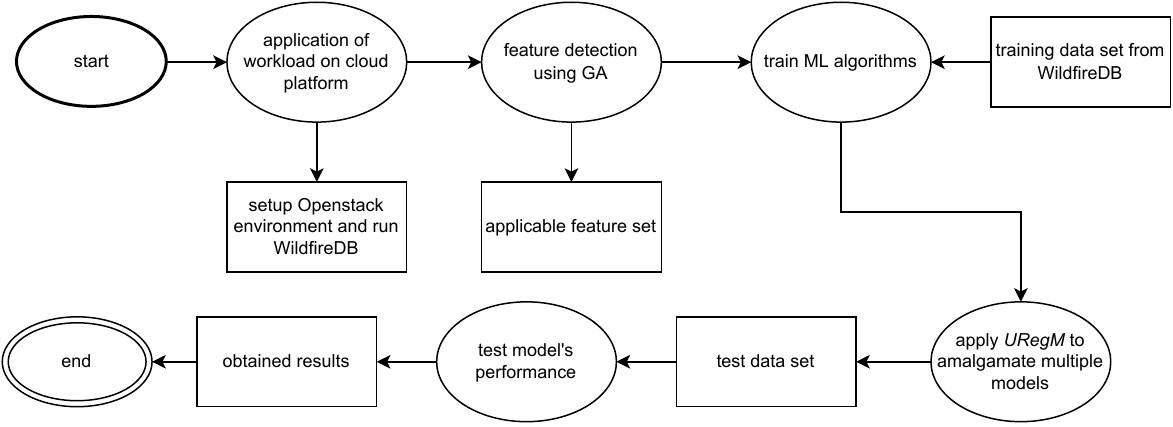}
    \caption{Workflow of the machine learning procedure}
    \label{mlwork}
\end{figure*}

Also, software engineers are placing time and effort to remove code smells in cloud platforms. As a result, predicting the change in resource requirement of the cloud after refactoring code smells is important to automatically and correctly provision future resource requirements and refactoring activities \cite{KRETSOU2021110892}. It will be beneficial from both the technical and financial perspectives of the cloud service provider. The ML approach used to predict CPU and memory usage in cloud for different workloads is called \textit{Unified Regression Model} (\textit{URegM}) based prediction. The various methods, classes, and data in cloud are dependent on each other. The resource consumption of a particular component before and after elimination of code smells may be caused by the other connected components. This potential dependency is modeled using regression techniques and the flow of the proposed model is shown in Figure \ref{mlwork}. We compare our obtained results to the existing individual models and \textit{REAP} \cite{kaur2019intelligent} approach to analyze performance.

\subsection{Feature Selection}

The important components of our Machine Learning (ML) prediction for resource usage are the population, feature selection, and ML algorithm. Feature selection is an important task of ML activity in various research. We use genetic algorithm (GA) for feature selection, which is a greedy approach that selects the best features based on produced output. Feature selection using GA provides the constraints and inputs of software which has significant impact on the resource consumption of the software. We provide the procedure of GA in Algorithm \ref{algortihm}.

Initially, random list of features is selected to form the population. Next the correctness of those features is determined by their fitness values. Acceptable fitness score range from \textit{76.0 to 89.0} \cite{verdecchia2018empirical}. The fitness values are scaled in an acceptable range which is called expectation scores. These scores are used to select the features for prediction of resource consumption. The features which get low scores will be removed and can be used together with the next set of population. New set of features can be obtained by conducting mutation and crossover on the correct population. Mutation and crossover can be done individually or together. The process can be repeated and if the current set of population of features outperforms the earlier set, then the current set of features will replace the previous set of features.

\begin{figure*}
    \centering
    \setlength{\belowcaptionskip}{-12pt}
        \includegraphics[width=\textwidth]{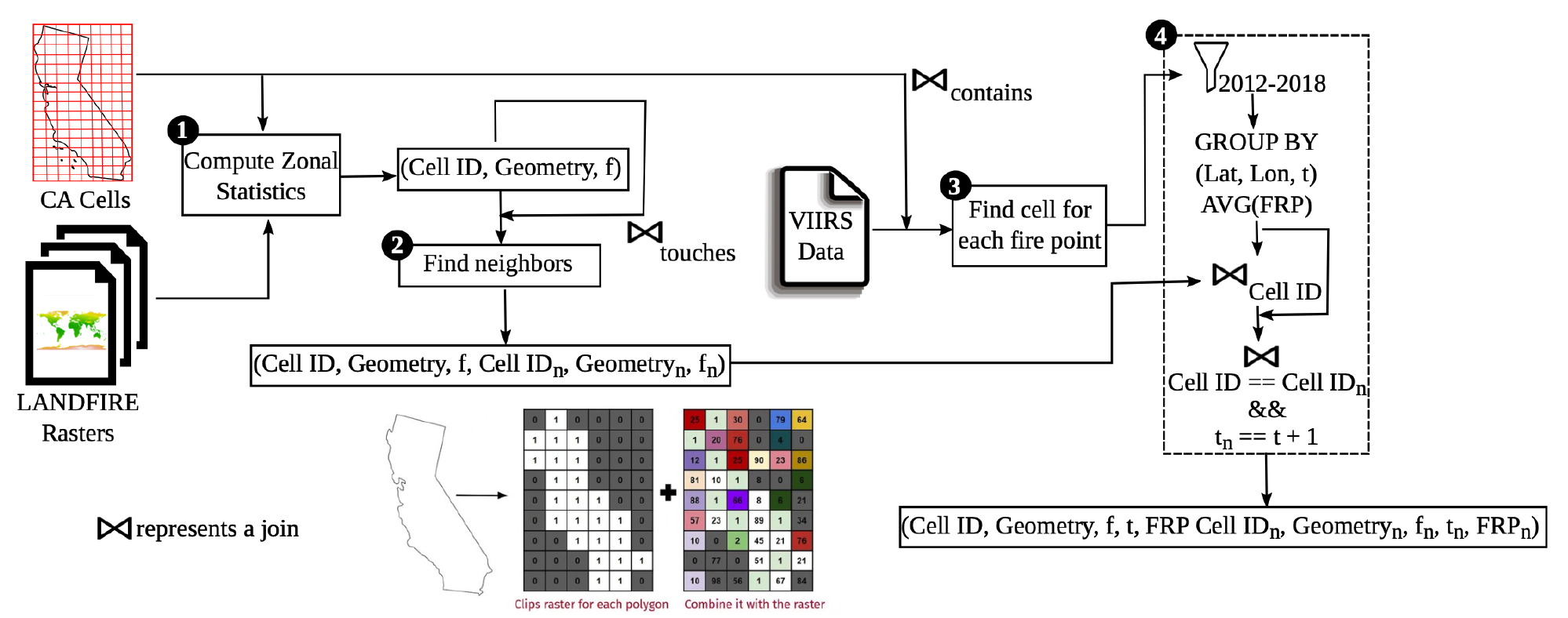}
    \caption{WildfireDB workload description \cite{singla2020wildfiredb}.}
    \label{cybershake}
\end{figure*}

\subsection{Workload description}

\textit{WildfireDB} is a tool for modeling fire occurrence and spread by analyzing data about locations, sizes, vegetation, field, and other topographic features \cite{singla2020wildfiredb}.  The architecture of the tool is shown in Figure \ref{cybershake}. The dataset includes \textit{2,367,209} data points of California wildfires. Each data point corresponds to a specific \textit{375m X 375m} polygon area at a given time, and it also records the condition of the neighboring cells at that time frame. The relevant covariates are also present for analysis. Similar statistics are provided regarding topographic information of vegetation, fuel, etc which includes the maximum, minimum, median, sum, mode, count, and mean values. \textit{Firesim} is incorporated with the database to simulate wildfire situations for climate change ecological analysis. Total simulation run was conducted for \textit{5001.68} seconds. 

Climate change ecologists face significant challenges in analyzing wildfire data because fire occurrence and covariates are available in various frameworks. Amalgamating the different frameworks and analyzing them require very high computation power. Secondly, the raster data in this dataset have more that two million data points for the state of California alone. Mining large scale feature data is a significant bottleneck in terms of computational resources. Third, the large size of data further complicates the resources required to combine them for analysis. Combining data sources from various frameworks require conversion of the data into a uniform representation. This conversion is computationally expensive and due to being two-dimensional data, the size will increase by 4 times when combined to a single uniform format. The computational complexity of combining the dataset of raster and vector data in\textit{WildfireDB} is {\[  O(np^2. c . r)\]} Here \textit{np} is the number of polygons, and \textit{c} and \textit{r} are the number of rows and columns in the dataset. 

\textit{WildfireDB} is designed to conduct this data fusion activity followed by analysis in a fully decentralized approach. To ensure that computer scientists can define systems which can make this process more resource friendly, \textit{WildfireDB} provides the distributed algorithms and data structures which can be run to stress the resources of cloud computing environment to analyze resource bottlenecks on the cloud processes. The simulation models which can be run to amalgamate the vector and raster data for generating the computationally expensive 4-dimensional dataset has been provided. At the same time it provides an algorithm which forms two parallel data sets to be used at run time. Those operations stress the exascale supercomputers like \textit{Frontera} of \textit{Texas Advanced Computing Center (TACC)} and \textit{Blue Waters} of \textit{National Science Foundation (NSF)}. Hence, the workload is substantial to test the impact on cloud resources.

\subsection{URegM framework}

The ML approach used to predict CPU and memory usage in cloud for different workloads is called \textit{URegM}. The various methods, classes, and data in cloud are dependent on each other. The resource consumption of a particular component before and after elimination of code smells may be caused by the other connected components. This potential dependency is modeled using regression techniques. The regression analysis helps us to detect change in CPU and memory usage for each clock cycle on the same workload for varying data sizes. Initially the workload is applied on vm instances launched in \textit{Openstack} and after a number of experimentation with the workload data, a regression based analysis of CPU and memory usage is obtained. Also, the null values in the dataset are removed as part of cleaning the dataset. Next, based on the obtained values of resource usage, GA is applied to select the features in the software code which affect the resource consumption. 

\begin{figure*}
    \centering
    \setlength{\belowcaptionskip}{-12pt}
        \includegraphics[width=1.0\textwidth]{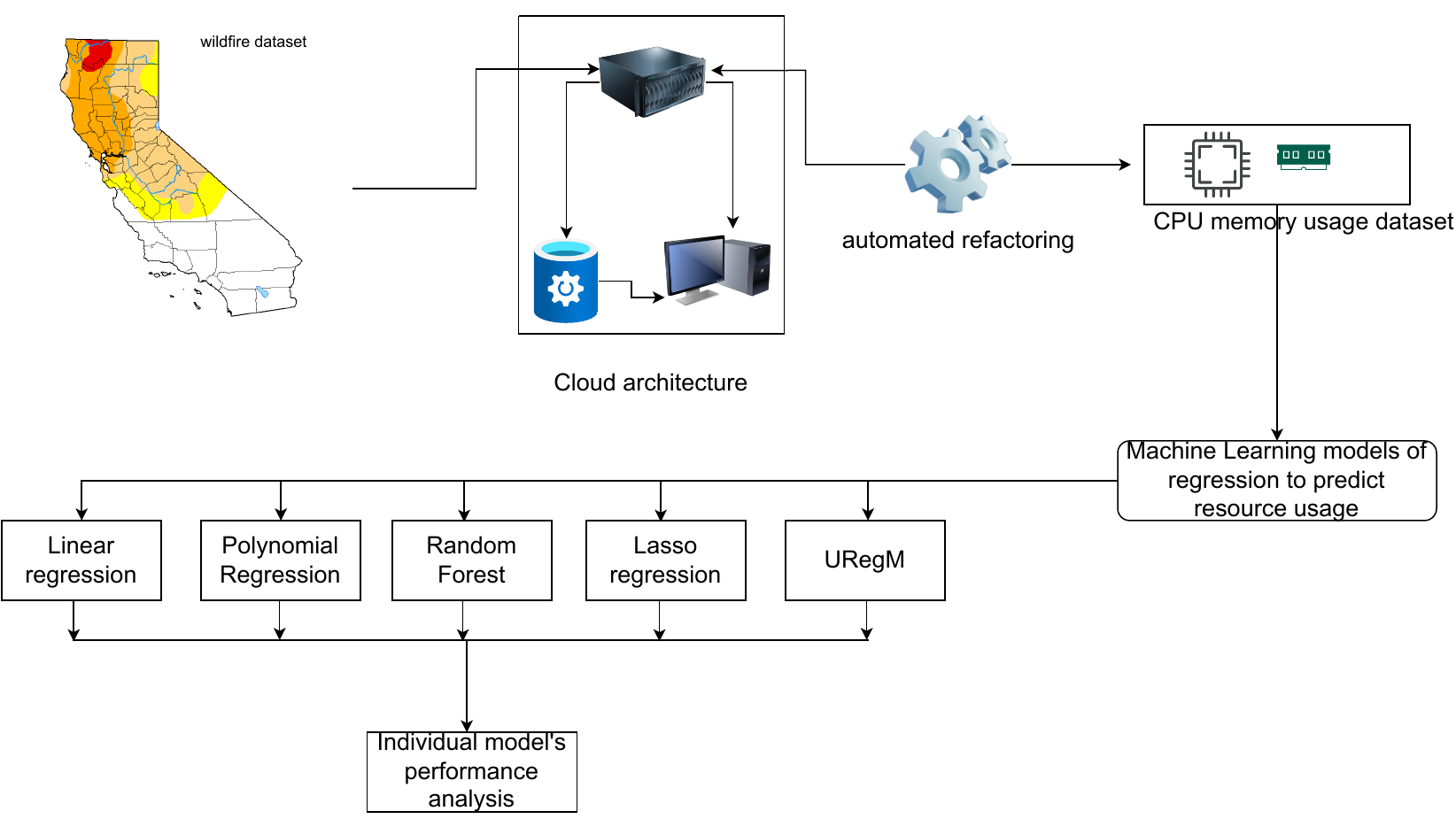}
    \caption{Application of workload and \textit{URegM} approach in Openstack.}
    \label{mlappr}
\end{figure*}

At the same time, only the components of the software where code smells are found and refactored are considered in the dataset, thereby significantly reducing the overall size of the dataset and removing redundant code blocks. Next, the ML algorithms are applied on the final dataset and regression models are trained. We divide the dataset into two halves. One half which contains 80\% of the data is used for training and 20\% of the data is kept in other half which is used for testing. Algorithm \ref{algortihm} is provided which explains the ML approach. The model is trained on the training dataset and tested on the test dataset, and the metrics such as \textit{mse, rmse, accuracy} and \textit{execution time} are used to test its performance. The training dataset is obtained from the benchmarking study where code smells in 31 open source software are refactored using automated tools and the change in resource consumption is monitored before and after refactoring for each method of the software where refactoring was conducted \cite{imran2021impact}. Pre-specified workloads are used for the benchmark study. The researchers establish a benchmark where individual types of code smells are detected and refactored in each software, followed by an analysis of CPU and memory consumption impact. Afterward, they conducted a batch refactoring of smells and analyze their collective impact on resource usage. We use the dataset of 31 open source software to train \textit{URegM}. Finally, the performance of the four ML models on the dataset together with the proposed \textit{URegM} approach are recorded on cloud environment.

In Algorithm \ref{algortihm}, $BSC$ is a variable which represent the accuracy of the various combination of the four individual models, $BMSc$ represent the combination of models which has highest accuracy, $models$ is an array which lists all the models used for making the combination. $DS$ contains the results of the prediction dataset which is applied to the various models for the training operation, datasets from \textit{0 to n-1} is considered, which is used and $Train$ contains the combination of regression models which is produced from the four individual regression approaches. Next, the $Test$ process assembles the models and compares the accuracy. In case the calculated accuracy is better than the $BSC$, then the value of $BSC$ is overwritten and the combination of models which gave the new accuracy is used to replace the existing set of models in $BMSc$. 

\begin{algorithm}[hbt!]
\caption{URegM algorithm}\label{alg:cap}
\begin{algorithmic}
\Require $BSc \geq 0$
\State $BMSc = NULL$
\State $models[i] \gets [LiR, PR, LR, RF, URegM]$
\State $DS \gets predictionSet[]$
\State $Main \gets predictionSet[1]$
\While{$N \neq 0$}
\While{x in 1...n}
    \State $Train \gets models[0..n-1] \times X$
    \State $Test \gets \models[n] $
    \State URegM[Test]
    \If{$MSc$ is URegM == predictionSet[0,1]*0.1}
    \If{$BMSc < MSc$}
    \State $BMSc \gets MSc$
    \State $BSc \gets DS$
\EndIf
\EndIf
\EndWhile
\EndWhile\\
    \Return $BMSc\ and\ URegM[results]$
\end{algorithmic}
\label{algortihm}
\end{algorithm}

\textit{URegM} combines prediction results from various algorithms to obtain overall better results and get stronger predictions. It is a supervised learning approach which can be trained and improved. Unifying the various models through \textit{URegM} will yield better results if the individual models are different from each other in terms of algorithmic design and data analysis. The ML procedure is shown in Figure \ref{mlappr}. The input workload is the \textit{WildFireDB} which consists of 268 GB of data and multiple files with various operations which are executed in cloud. Overall the CPU usage, memory usage, and file sizes are shown in Table \ref{vminstance}. The experimentation is conducted with various number of tasks executed by the workload on the cloud and the CPU and memory usages are predicted. 

A positive correlation is determined between the number of tasks and the CPU usage, on the other hand, positive correlation is determined between the size of data and memory usage.  For example, CPU usage for 500 tasks came out to be 3.6\% whereas for 1500 tasks the usage incremented to 4.2\%. It increased to 5.0\% for 2000 tasks, then again jumped to 7.8\% for 4000 tasks as observed. It must be mentioned that CPU and memory usage is monitored for the methods and classes which are impacted due to batch refactoring, the resource usage by the other parts of the code are ignored within the scope of this research. It can be determined that CPU resource usage is dependent on the number of tasks. After refactoring code smell, memory usage for 500, 1500, 2000, and 4000 tasks came out to be 2.9\%, 3.8\%, 4.3\%, and 7.4\% respectively. 

\begin{table}
\centering
\begin{tabular}{ l c c c c } 
 \hline
 \textbf{VMid} & \textbf{vCPU} & \textbf{RAM (GB)} & \textbf{Disk (GB)} & \textbf{OS} \\ \hline 
 1 & 1 & 0.5 & 50 & Ubuntu 18.04\\
 \hline 
 2 & 1 & 1 & 256 & CentOS 7\\
 \hline 
 3 & 2 & 2 & 500 & Ubuntu 18.04\\
 \hline 
4 & 2 & 4 & 500 & CentOS 7\\
 \hline 
5 & 4 & 4 & 500 & Ubuntu 18.04\\
 \hline
6 & 4 & 6 & 500 & CentOS 7\\
 \hline
\end{tabular}
\caption{Description of virtual machines launched in OpenStack}
\label{vminstance}
\end{table}
\section{Experimental setup}
\label{experiment}
In this section, the experimental process of \textit{URegM} approach is discussed. Initially, the experimental environment is highlighted, followed by discussion on the obtained results.

\subsubsection{Experimental environment}

We run multiple tasks of \textit{WildFireDB} in cloud virtual machines to generate the workload and test the resource consumption before and after refactoring. \textit{OpenStack} cloud platform is used to run the workload of \textit{WildFireDB}. \textit{OpenStack} is setup in \textit{HP Proliant DL380P} server which had 8 cores and 32 GB RAM. We launched 6 virtual machine instances using kernel virtual machine (kvm) based virtualisation. The 6 virtual machine instances were setup and the workload was executed in those. The virtual machine instances were heterogeneous. We justify the obtained results of resource consumption prediction by running the workload in cloud environment.

\begin{table}
\centering
\begin{tabular}{ l c c } 
 \hline
 \textbf{Tasks} & \textbf{Predicted \% of CPU} & \textbf{Actual \% of CPU} \\ \hline 
 500 & 3.6 & 3.8 \\
 \hline 
 1000 & 4.0 & 4.1\\
 \hline 
 1500 & 4.2 & 4.6\\
 \hline 
2000 & 5 & 5.3\\
 \hline 
2500 & 5.9 & 5.9\\
 \hline
3000 & 6.6 & 6.7\\
 \hline
 3500 & 7.3 & 7.6\\
 \hline
 4000 & 7.8 & 8.2\\
 \hline
\end{tabular}
\caption{Prediction of CPU usage}
\label{vminstancecpu}
\end{table}

\begin{table}
\centering
\begin{tabular}{ l c c } 
 \hline
 \textbf{Tasks} & \textbf{Predicted \% of memory} & \textbf{Actual \% of memory} \\ \hline 
 500 & 2.9 & 3.4 \\
 \hline 
 1000 & 3.4 & 3.7\\
 \hline 
 1500 & 3.8 & 3.10\\
 \hline 
2000 & 4.3 & 4.5\\
 \hline 
2500 & 4.9 & 4.16\\
 \hline
3000 & 6.2 & 6.9\\
 \hline
 3500 & 6.12 & 6.28\\
 \hline
 4000 & 7.4 & 7.9\\
 \hline
\end{tabular}
\caption{Prediction of memory usage}
\label{vminstancememory}
\end{table}

The 6 virtual machines are described in Table \ref{vminstance}. It is seen from the properties of the virtual machines that they have different configurations. This simulates the distributed environment in which the applications run in cloud. This also helps us to generate a real life scenario and so we can proceed to obtain the resource usage of cloud which will reflect real life consumption of CPU and memory.
\begin{table*}
\centering
\begin{tabular}{l c c c c c c c c c c}
\hline
\textbf{Metrics} & \textbf{LiR} & \textbf{PR} & \textbf{LR} & \textbf{RF} & \textbf{REAP} & \textbf{URegM}\\ \hline
mse & 1.47 & 0.72 & 0.56 & 0.40 & 0.27 & 0.21 \\ \hline
rmse & 1.66 & 0.94 & 0.74 & 0.60 & 0.37 & 0.29 \\ \hline 
accuracy (\%) & 86.70 &  90.60 & 88.91 & 93.31 & 95.41 & 96.22 \\ \hline
time (s) & 3.60 & 1.54 & 1.67 & 1.89 & 0.48 & 0.33 \\ \hline 
\end{tabular}
\caption{Performance of the ML models. (LiR:=Linear Regression; PR:= Polynomial Regression; LR:= Lasso Regression; RF:= Random Forest; REAP:= Regressive Ensemble Approach for Prediction); URegM (Unified Regressive Model)}
\label{modelanalysis}
\end{table*}
We aim to predict the CPU and memory usage change due to refactoring code smells for low, medium, and high workload. This is because the number of tasks performed by the workload may impact the resource usage. Table \ref{vminstancecpu} highlights the predicted CPU usage change of the different VMs with various workloads before and after eliminating code smells. The various workloads applied are \textit{firesim\_500, firesim\_1000, firesim\_1500, firesim\_2000, firesim\_2500, firesim\_3000, firesim\_ 3500,} and \textit{firesim\_4000} which are deployed in the 6 virtual machine instances starting \textit{VMid\_1} to \textit{VMid\_6}. This covers all possible situations in which the resource usage prediction models may face in real life cloud environment. Each workload was executed 6 times and the average CPU and memory usage was calculated. In Table \ref{vminstancememory} we identify the memory usage difference before and after refactoring code smells of the class files where the code smells were refactored. We proceeded on analyzing the impact of refactoring 5 code smell types namely \textit{god class, god method, cyclic dependency, long parameter,} and \textit{spaghetti code}.

\subsubsection{Results Analysis}
The outcome of extensive experimental analysis of the proposed approach is described in this section.  The unified model is applied so that it can combine the strengths of the individual models and obtain effective performance of predicting resource usage changes due to code smell refactoring. Initially, we analyze the performance of the individual regression models. Next, we assess the performance of the ensemble model. The experimental results are shown in Table \ref{modelanalysis}. Among the various regression models, we selected these 4 models to form the unified model based on their effectiveness and ease of implementation using the Algorithm \ref{algortihm} shown in this chapter. The \textit{REAP} model is an ensembled model for prediction which is also executed on the dataset to compare the performance of our proposed framework.

\iffalse 
\begin{figure*}
    \centering
    \setlength{\belowcaptionskip}{-12pt}
        \includegraphics[width=1.0\textwidth, height = 250]{barplot2.png}
    \caption{Performance analysis of different models.}
    \label{mlperformanceanalysis}
\end{figure*}
\fi
It is seen that individual models do not perform to the extent of the unified approach. The performance of the models depend on the training dataset which is used in the \textit{WildFireDB} workload. With the change in dataset, the selected evaluation metrics may show different results. It is seen that individual model may have better error rate than unified models, however, they have worse accuracy and execution time. To obtain the best performance, the individual algorithms are combined together to form the unified model. The accuracy of the proposed algorithm is seen to be highest at 96.22\% and it has an average execution time of 0.33 seconds. The overall performance in terms of accuracy is increased by 1.8\% and the execution time is decreased by 14.2\%. Hence it can be concluded that for this task, \textit{URegM} outperforms the individual regression models. The results of the prediction technique is promising and it shows that estimation of resource consumption due to refactoring code smells even before those smells are refactored will help the software engineers decide on the refactoring. The approach can be applied for other workloads besides the \textit{WildFireDB}. As a result, this study helps the correct resource provisioning and decisions for large scientific applications in cloud computing environment.

\section{Conclusion}
\label{conclusion}

In this paper, we proposed a method to predict change in computing resource requirements due to the refactoring of cloud source code. High accuracy of CPU and memory usage change prediction due to refactoring code smells was obtained by \textit{URegM} which used \textit{GA} to remove noise and irrelevant information from the dataset. A cloud computing environment was setup with a real-life scientific application to analyze the performance of the intelligent \textit{URegM} algorithm. It was seen that although the individual algorithms were effective, \textit{URegM} outperformed those by improved error rate and faster execution rate. Also, accuracy was improved by 2.3\% using \textit{URegM}, whereas the execution time was reduced by 10.9\%. The results showed that \textit{URegM} had better accuracy and execution time compared to both the individual learning automata-based approaches and \textit{REAP} model.

In the future, an incremental learning paradigm may be added to enable \textit{URegM} to accommodate the different types of code smells that are refactored by different auto-refactoring tools. This will make the prediction approach more adaptive. Also, we will study the relationship between refactoring code smells and its impact on software complexity \cite{shehab2015accumulated}. We will also aim to test the performance of the proposed model on other cloud platforms.
\begin{acks}
This project is in part sponsored by the National Science Foundation (NSF) under award numbers OAC-1724898, OAC-1842054 and CCF-2007829.
\end{acks}
\bibliography{esse}
\end{document}